# Effect of phonon dispersion on thermal conduction across Si/Ge interfaces[1]

Dhruv Singh      Jayathi Y. Murthy*      Timothy S. Fisher

School of Mechanical Engineering and Birck Nanotechnology Center,

Purdue University, West Lafayette, IN -47907, USA

## ABSTRACT

We report finite-volume simulations of the phonon Boltzmann transport equation (BTE) for heat conduction across the heterogeneous interfaces in SiGe superlattices. The diffuse mismatch model incorporating phonon dispersion and polarization is implemented over a wide range of Knudsen numbers. The results indicate that the thermal conductivity of a Si/Ge superlattice is much lower than that of the constitutive bulk materials for superlattice periods in the submicron regime. We report results for effective thermal conductivity of various material volume fractions and superlattice periods. Details of the non-equilibrium energy exchange between optical and acoustic phonons that originate from the mismatch of phonon spectra in silicon and germanium are delineated for the first time. Conditions are identified for which this effect can produce significantly more thermal resistance than that due to boundary scattering of phonons.

---

# I. INTRODUCTION

During the past decade, there has been growing interest in developing nanostructured materials such as composites and superlattices for use in thermoelectrics, thermal interface materials and in macroelectronics [1-5]. It has long been known that the thermal and electrical transport properties of isolated nanostructures such as thin films, nanowires and nanotubes often exhibit significant deviations from their bulk counterparts [6-8]. Concomitantly, a developing body of work has enabled deeper understanding of thermal transport in these materials, and the prediction of their effective thermal conductivity [9-12]. It is therefore important to understand how nanoscale transport phenomena influence the thermal behavior of superlattice and nanocomposite materials.

Three types of nanoscale physical phenomena may manifest themselves when considering thermal transport in superlattice structures: interface and boundary scattering, wave effects, and transport across heterogeneous interfaces. Enhanced interface and boundary scattering causes a reduction in the mean free path of phonons in nanostructured materials. Experimentally observed reductions in the thermal conductivity of submicron wires and thin films have been reported in [6, 8], and models based on the phonon Boltzmann transport equation (BTE) have also captured this phenomenon [9-12]. When nanostructuring reaches extremely small dimensions (*e.g.*, a few atoms) wave effects must be considered [13-15]. Band folding in superlattices increases with increasing period length and leads to a reduction in the average phonon group velocity [16]. This behavior is particularly prevalent for period lengths starting from a small number of atomic layers and would be expected to lead to a decrease in superlattice thermal conductivity as the period length increases in this regime. This prediction contradicts particle theory, which predicts a monotonic increase in thermal conductivity as the superlattice period length is increased due to decreasing interfacial density. As pointed out by Simkin and Mahan [16], the transition from wave to particle behavior in phonon transport leads to a minimum in thermal conductivity with respect to superlattice period. Such a trend has been observed experimentally in the $Bi_2Te_3$/ $Sb_2Te_3$ superlattices, which exhibit a minimum in thermal conductivity for a superlattice period near 50 Å [18]. Non-equilibrium MD simulations have also revealed a minimum in the thermal conductivity as superlattice period decreases [17]. These simulations however show that the trend quickly disappears with a mismatch in the lattice constant of the two materials. MD Simulations in [54] also show that even a small mixing of interfacial species in Si/Ge superlattices leads to a crossover to incoherent phonon transport wherein thermal conductivity increases as period length increases. For superlattices, transport of heat across heterogeneous material interfaces plays a critical role. The acoustic and diffuse mismatch models (AMM and DMM) [19, 20] have formed the mainstay of most published analyses, though their applicability has been called into question by experimental data [1]. More recently, the phonon wave packet technique [21-23] and atomistic Green's functions (AGF) [24] have been used to determine interface transmission functions.

A number of theoretical studies have appeared in the recent literature which attempt to quantify the influence of wave effects on transport in superlattice structures. Ren and Dow [25] suggested the existence of mini-Umklapp phonon scattering processes associated with the Brillouin zone corresponding to the superlattice period length. Using a phenomenological model, they qualitatively demonstrated that the strength of mini-Umklapp scattering increases as period length is made smaller and is directly influenced by the mass difference between the constituent atoms. More recently, Broido and Reinecke [26] and Ward and Broido [27] have taken a direct approach based on empirical interatomic potentials in calculating the intrinsic lattice thermal conductivity of Si/Ge and GaAs/AlAs superlattices. Their calculations show that the reduction in phonon group velocity with increasing mass ratio is primarily responsible for decrease in the superlattice thermal conductivity, consistent with other studies based on lattice dynamics. However, they also show that changes in phonon dispersion lead to a reduction in the phase space for three-phonon scattering events, thus increasing the average phonon lifetime. This is contrary to the view that the scattering rate increases due to zone folding in superlattices [1, 25]. The increase in phonon lifetime due to this effect is still not sufficient to overcome the reduction caused by changes in group velocity, with the exception of 1x1 superlattices.

Concurrently, a strong experimental effort has been led by several groups to understand phonon transport and thermal conductance of solid-solid interfaces. Simplified acoustic mismatch and diffuse mismatch models explained initial data on heat transfer between dissimilar solids [19,20]. Review articles by Cahill *et al.* [1, 28] summarize developments in measurements and modeling of heat transfer across solid interfaces. Ref [1] discusses the key experimental results on Si/Ge, Si/SiGe, GaAs/AlAs, $Bi_2Te_3$/$Sb_2Te_3$ superlattices and alloys. Lee *et al.* [29] showed that at superlattice periods less than 10 nm, the effective thermal conductivity of Si/Ge superlattices decreases as the superlattice period decreases, pointing to the dominance of interface scattering. This is contrary to the theoretical expectation that thermal conductivity should increase due to wave interference as superlattice period becomes very small [16, 26, 27]. Their results also showed that at large superlattice period lengths, thermal conductivity was significantly lower than that at small period lengths and almost independent of period length, a behavior that was attributed to the poor crystal quality of the larger period length superlattices. However, the corresponding interfacial thermal resistance implied by the effective thermal conductivity is much lower than that calculated by the popular diffuse mismatch model (DMM), indicating that DMM assumptions are inappropriate at such epitaxial interfaces [29]. Calculations of thermal resistance in the present work using a full dispersion DMM show that the discrepancy with respect to experiments can be as high as 3-5 times at 300K for short-period superlattices. Due to a high degree of acoustic mismatch between Si/Ge, the acoustic mismatch model (AMM) predicts an even higher value of interfacial thermal resistance [24, 30] than the DMM. Theoretical modeling results [30] suggest that partially specular interface scattering can explain the experimentally observed values only when inelastic effects in phonon transmissivity are taken into account.

Other experimental studies also point to the importance of interface scattering and acoustic mismatch in superlattice structures. Experimental studies have found that Si/Ge superlattices can exhibit lower thermal conductivity than the corresponding $Si_\varphi Ge_{1-\varphi}$ alloys [29, 31] and a strong dependence on doping [31], suggesting interface and impurity scattering effects are important. Systematic studies by Huxtable *et al.* [32] on Si/SiGe (Si/alloy) and SiGe/SiGe (alloy/alloy) superlattices were performed to unravel the effects of acoustic mismatch, alloy scattering and period length. They found a monotonic decrease in thermal conductivity with decreasing superlattice period, eliminating the possibility of phonon bandgap formation and band folding as being the primary mechanism in impeding thermal transport. While Si/alloy superlattices exhibit a strong dependence of thermal conductivity on period length, the thermal conductivity of alloy/alloy superlattices is found to be almost independent of the period length. These results point to acoustic mismatch at interfaces governing thermal transport in Si/SiGe superlattices and alloy scattering in SiGe/SiGe superlattices. Similar studies on Si/SiGe superlattice nanowires were performed by Li *et al.* [33]; their results indicate a lower thermal conductivity than the superlattice films due to additional boundary scattering of phonons. A theoretical study by Dames and Chen [34] on similar nanowire superlattices seems to suggest that incoherent particle-like behavior can explain the experimental trends. The existence of a definite minimum superlattice thermal conductivity has thus far not been observed in experiments on Si/Ge or SiGe alloy superlattices and may have been masked by strain effects.

These experimental studies suggest that particle theories would be effective in modeling transport in heterogeneous structures such as superlattices. The Boltzmann transport equation (BTE) under the relaxation time approximation with a model for interface transmissivity was used by Chen and coworkers [30,34-37] to predict transport in superlattice structures. They employed a gray approximation for thermal transport in the constituent materials, with models such as the DMM for transmission across multi-material interfaces. Frequency-dependent effects were approximated through the use of an integrated intensity. These models have also addressed the effect of diffuse interface scattering on the thermal conductivity of planar and nanowire superlattice structures by employing interface specularity as an empirically adjusted parameter.

In this paper, we investigate the transport processes involved in frequency-dependent phonon transport across a single Si/Ge interface and in Si/Ge superlattices. The intent is to examine the role of phonon dispersion and polarization in determining overall thermal conductance as well as detailed transport characteristics for heat transfer across heterogeneous interfaces. Recently it has been shown that departure from bulk dispersion due to phonon confinement in silicon is limited to sizes below 5 nm for wires and 10 nm for thin films [38]. In keeping with these calculations and observed experimental trends, we employ an incoherent particle model based on the Boltzmann transport equation (BTE). Bulk modes for Si and Ge are assumed. This approach is appropriate in the transition regime for which both bulk and interface scattering play a strong role in determining effective thermal conductivity, and when wave effects are negligible. A frequency-dependent diffuse mismatch model (DMM) is employed. Our choice of diffuse interface scattering is justified by recently published simulations of phonon scattering across rough fcc lattices using the atomistic Green's function method [24]. The results show that specular or partially specular phonon reflection is mostly limited to very long wavelength phonons while medium and short wavelength phonons undergo complete diffuse scattering. We employ empirical frequency-dependent bulk scattering rates. Our computations indicate that the selective filtering of phonon modes at heterogeneous interfaces plays a significant role in decreasing the thermal conductance of superlattice structures, an effect not captured by gray BTE models. Details of the filtering mechanism and its interaction with bulk scattering are also revealed.

## II. THEORY AND MODELING

Beyond the realm of quantum confinement, bulk dispersion curves and phonon group velocities may be defined and the Boltzmann transport equation may be used to describe phonon heat conduction. The primary difference between phonon heat conduction in homogeneous nanostructures and nanostructured composites arises from phonon scattering at the heterogeneous interfaces. We describe thermal transport in the semiconductor using the multiple-band, single-mode relaxation time BTE. While a gray model captures the essential mechanism of thermal transport in the cross-plane direction of superlattices, it ignores the details of the phonon spectra in individual materials [30].

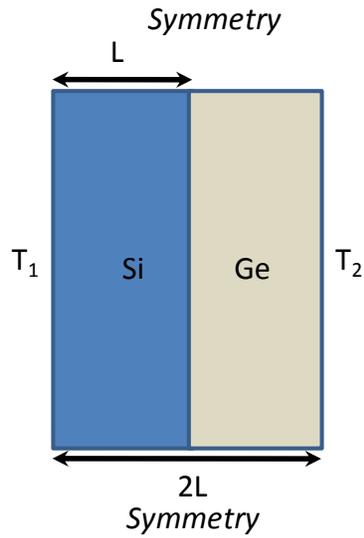

Fig. 1(a): Domain for single interface simulation

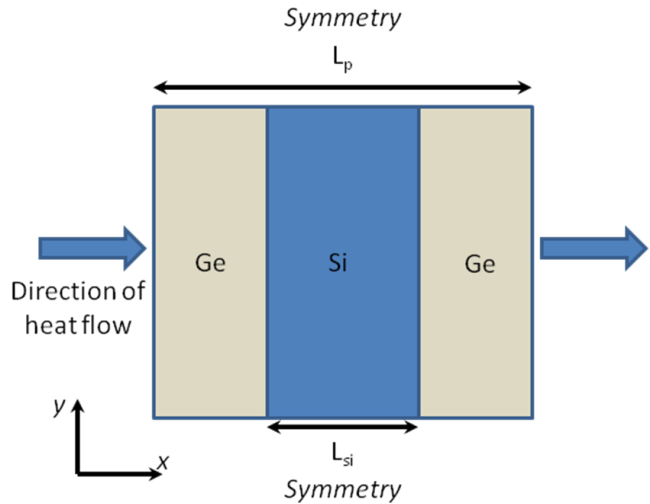

Fig. 1(b): Periodic unit cell of the superlattice

Two types of problems are considered here. The first, involving transport across a single Si/Ge interface, is shown in Fig. 1(a). The second involves the simulation of thermal conductivity in an infinitely long Si/Ge superlattice; the corresponding periodic module is shown in Fig. 1(b). Here, layers of Si with width $L_{si}$ and those of Ge are alternately arranged to form a superlattice with a period $L_p$. For finite-sized superlattices, a number $n$ of these modules would be stacked end-to-end to form the overall computational domain. However, to simulate a superlattice of infinite extent, a single module, with suitable periodic boundary conditions, is considered, as described below.

The Boltzmann transport equation for the energy density of phonons under a gray assumption [39] and the relaxation time approximation may be expressed as:

$$\nabla \cdot (v_g \mathbf{s} e'') = \frac{e^0 - e''}{\tau} \quad (1)$$

Here, $e''(\mathbf{r},\mathbf{s})$ is the net energy density (J/m$^3$sr) of all phonon groups in the solid with group velocity $v_g$ and relaxation time $\tau$ at position $\mathbf{r}$ and in direction $\mathbf{s}$. The quantity $e^0(\mathbf{r})$ represents the angular average of $e''(\mathbf{r},\mathbf{s})$ over all directions $\mathbf{s}$ at a given position $\mathbf{r}$. The gray BTE has been used to successfully probe ballistic-diffusive transitions in the cross-plane direction of superlattices and nanocomposites [30, 37] and is useful in understanding the mechanisms of phonon transport and the classical size effect, at least in a qualitative fashion. However, it ignores polarization and dispersion effects that are central to determining transport across heterogeneous interfaces.

The gray BTE model described in prior work may be extended to include frequency- and polarization-dependent relaxation times and velocities, as has been done for the equation of phonon radiative transfer [39, 40]. Assuming Brillouin zone isotropy, the band-wise BTE can be expressed as

$$\nabla \cdot (v_{\omega,p} \mathbf{s} e''_{\omega,p}) = \frac{e^0_{\omega,p} - e''_{\omega,p}}{\tau_{\omega,p}}$$

$$e''_{\omega,p} = \frac{1}{4\pi} \int_{\Delta\omega} f \hbar \omega D_p(\omega) d\omega \quad (2)$$

$$e^0_{\omega,p} = \frac{1}{4\pi} \int_{\Delta\omega} \frac{\hbar\omega}{e^{\hbar\omega/k_b T_L} - 1} D_p(\omega) d\omega = \frac{C_{\omega,p}}{4\pi} T_L$$

where $e''_{\omega,p}(\mathbf{r},\mathbf{s})$ is the energy density of phonons of polarization $p$ in the given material within a frequency band $\Delta\omega$ about frequency $\omega$ at the position $\mathbf{r}$ travelling along the wave vector direction $\mathbf{s}$ with a velocity $v_{\omega,p}$. $D_{pi}(\omega)$ represents the density of states for phonons of polarization p at frequency $\omega$. The quantity $e^0_{\omega,p}$ represents the mean energy density to which the phonons within this band relax.

The quantity $C_{\omega,p}$ is the volumetric specific heat associated with the discrete frequency band and polarization $p$ in the solid and $T_L$ is the lattice temperature.

Because scattering is purely re-distributive energetically, the net scattering term on the right side of Eq. (2) must integrate to zero over the Brillouin zone. An equivalent lattice temperature $T_L$ corresponding to this requirement is defined at each spatial position as

$$\sum_p \int_{\omega=0}^{\omega_{max}} \frac{1}{\tau_{\omega,p}} \frac{\hbar\omega}{e^{\hbar\omega/k_b T_L}-1} D_p(\omega) d\omega = \sum_{p,\omega^*} \int \frac{e''_{\omega^*,p}}{\tau_{\omega^*,p}} d\Omega$$

$$\sum_{p,\omega^*} \frac{C_{\omega^*,p} T_L}{\tau_{\omega^*,p}} = \sum_{p,\omega^*} \int \frac{e''_{\omega^*,p}}{\tau_{\omega^*,p}} d\Omega \qquad (3)$$

where the sum over $\omega^*$ runs over all discrete frequency bands, and $\omega_{max}$ is the maximum frequency that phonons in the solid can support. Since different phonon groups interact at widely different rates with the lattice bath, the rate of energy exchange determines the lattice temperature. Once the lattice temperature $T_L$ is defined, the corresponding equilibrium energy $e^0_{\omega,p}$ to which phonons in the band $\omega^*$ relax may be calculated using Eq. (2). Though a 'single-mode relaxation time' $\tau_{\omega,p}$ such as that used in this formulation does not completely capture anharmonic interactions (*i.e.*, creation and annihilation of phonons according to the selection rules), the energy exchange between different phonon bands is indirectly included using the concept of lattice temperature, and not by explicitly satisfying phonon energy and quasi-momentum conservation rules as in prior work [41]. Equivalent descriptions have been used in to calculate thermal transport along the length of Si nanowires [12], and the model has been quantitatively validated [42] for LJ argon. It can be easily shown that under small temperature gradients, high acoustic thickness ($Kn \ll 1$ for all frequency bands) and small departures from equilibrium, the BTE in each frequency band tends to a diffusion equation expressed as,

$$\left(\frac{C_{\omega,p} v^2_{\omega,p} \tau_{\omega,p}}{3}\right) \nabla^2 T_{\omega,p} = \frac{C_{\omega,p}(T_L - T_{\omega,p})}{\tau_{\omega,p}} \qquad (4)$$

where $C_{\omega,p,i}$ is the volumetric specific heat of the phonons in the solud of polarization $p$ in a frequency band $\omega$. The diffusion coefficient $C_{\omega,p} v^2_{\omega,p} \tau_{\omega,p}/3$ may be interpreted as an effective thermal conductivity of phonons within a band, and $T_{\omega,p}$ represents the corresponding band temperature.

*Uniform-temperature boundaries*

In investigating transport across a single interface, we apply a given temperature boundary condition on the two ends of the domain (see Fig. 1a). For a boundary with given temperature $T=T_b$, the energy density of all wave vector directions entering the domain from the boundary ($\mathbf{s}\cdot\mathbf{n} \leq 0$) is given by:

$$e_{\omega,p}^{"} = e_{\omega,p}^{0} = \frac{C_{\omega,p} T_b}{4\pi} \tag{5}$$

Here, **n** is the outward-pointing normal from the domain. The volumetric specific heat $C_{\omega,p}$ is assumed to be constant because the temperature difference, $T_1$-$T_2$, across the domain is assumed small. For all directions outgoing from the domain, the following boundary condition is used.

$$\nabla e_{\omega,p}^{"} \cdot \mathbf{s} = 0 \tag{6}$$

*Symmetry boundaries*

The top and bottom boundaries of the domains in Figs. 1a and 1b are modeled as symmetry boundaries. Here, specular reflection at the boundaries is enforced. For phonons directions incoming to the domain ( $\mathbf{s} \cdot \mathbf{n} \leq 0$ ),

$$e_{\omega,p}^{"}(\mathbf{s},\mathbf{r}) = e_{\omega,p}^{"}(\mathbf{s_r},\mathbf{r}) \tag{7}$$

where $\mathbf{s}_r$ is the specular direction corresponding to **s** incoming to the boundary

$$\mathbf{s} = \mathbf{s_r} - (2\mathbf{s_r} \cdot \mathbf{n})\mathbf{n} \tag{8}$$

and **n** is the outward-pointing normal at the boundary.

*Periodic jump*

A periodic jump boundary condition is used to simulate the heat transfer in periodic structures in the bulk limit [37]. In this limit, the 'temperature' of each periodic module must fall by a constant amount in the direction of heat flow. Correspondingly, the energy densities at the unit cell boundaries perpendicular to the heat flow direction in Fig. 1b must be related by,

$$\begin{aligned} e_{\omega,p}^{"}(\mathbf{s},0,y) &= e_{\omega,p}^{"}(\mathbf{s},L_p,y) + \frac{1}{4\pi} C_{p,\omega} \Delta T_{drop} & \mathbf{s.t} > 0 \\ e_{\omega,p}^{"}(\mathbf{s},L_p,y) &= e_{\omega,p}^{"}(\mathbf{s},0,y) - \frac{1}{4\pi} C_{p,\omega} \Delta T_{drop} & \mathbf{s.t} < 0 \end{aligned} \tag{9}$$

where $\Delta T_{drop}$ signifies the temperature drop in the unit cell in the heat flow direction **t**. We note that this type of jump periodicity is only possible if the problem is linear, *i.e.*, if $C_{\omega,p}$ and $\tau_{\omega,p}$ are independent of lattice temperature.

*Interface between two solids*

Phonon transport across solid-solid interfaces is treated in a manner generally similar to that described previously [37] but including explicitly the effects of phonon dispersion and polarization. The dominant phonon wavelengths near room temperature are assumed to be much smaller than the inherent surface roughness, so that diffuse scattering at the interface can be assumed. Under these conditions the diffuse mismatch model (DMM) of Swartz and Pohl [20] is applicable and assumes that a phonon loses memory of its incident wave vector and polarization *p* upon encountering the interface. The only restriction imposed is that the scattering event is elastic, *i.e.*, the frequencies of incident and reflected phonons are identical. While the original DMM analysis in [20] does not include details of phonon dispersion, these are easily incorporated [43].

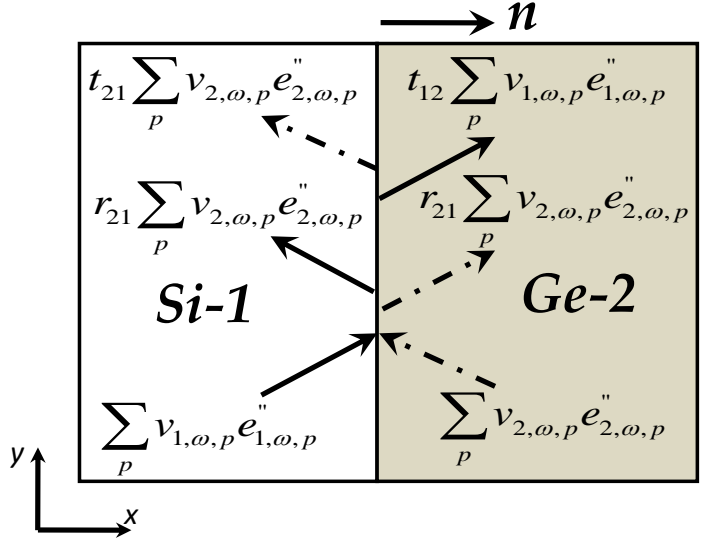

*Fig. 2: Schematic of phonon transport across a Si-Ge interface*

A schematic illustrating the interfacial energy balance is shown in Fig 2. Here, material 1 is Si and material 2 is Ge. The sum $\sum_p v_{1,\omega,p} e''_{1,\omega,p}$ represents the directional phonon energy flux carried by phonons in a frequency band $\omega$ (summed over all polarizations) incoming from side 1 to the interface. The transmissivity of the phonon energy density from medium 1 to medium 2 is denoted by $t_{12}$ and is a function of the phonon frequency $\omega$. The reflectivity of the interface to energy incoming from medium 1 is denoted by $r_{12}$, with $t_{12} = 1 - r_{12}$; $t_{21}$ and $r_{21}$ denote these quantities for transmission and reflection from medium 2. The DMM can be used to calculate a frequency dependent value of the transmissivity as [20, 43]

$$t_{12} = \frac{\sum_p v_{2,\omega,p} C_{2,\omega,p}}{\sum_p v_{1,\omega,p} C_{1,\omega,p} + \sum_p v_{2,\omega,p} C_{2,\omega,p}} \tag{10}$$

$$t_{21} = 1 - t_{12}; r_{12} = 1 - t_{12}$$

Because phonon scattering at the interface is assumed diffuse, a detailed energy balance leads to the following expression for the interface energy density as a function of phonon frequency and polarization,

$$e^{"}_{f,\omega,p}(\mathbf{s}.\mathbf{n} > 0) = \frac{1}{\pi} \frac{C_{2,\omega,p}}{\sum_p v_{2,\omega,p} C_{2,\omega,p}} \left( \begin{array}{l} r_{21} \int_{\mathbf{s}.\mathbf{n}<0} \sum_p v_{2,\omega,p} e^{"}_{2,\omega,p} \mathbf{s}.\mathbf{n} d\Omega + \\ t_{12} \int_{\mathbf{s}.\mathbf{n}>0} \sum_p v_{1,\omega,p} e^{"}_{1,\omega,p} \mathbf{s}.\mathbf{n} d\Omega \end{array} \right)$$

$$e^{"}_{f,\omega,p}(\mathbf{s}.\mathbf{n} < 0) = \frac{1}{\pi} \frac{C_{1,\omega,p}}{\sum_p v_{1,\omega,p} C_{1,\omega,p}} \left( \begin{array}{l} r_{12} \int_{\mathbf{s}.\mathbf{n}>0} \sum_p v_{1,\omega,p} e^{"}_{1,\omega,p} \mathbf{s}.\mathbf{n} d\Omega + \\ t_{21} \int_{\mathbf{s}.\mathbf{n}<0} \sum_p v_{2,\omega,p} e^{"}_{2,\omega,p} \mathbf{s}.\mathbf{n} d\Omega \end{array} \right)$$

(11)

where $e^{"}_{f,\omega,p}$ denotes the energy density of phonons of frequency $\omega$ and polarization $p$ at the interface, $e^{"}_{1,\omega,p}$, $e^{"}_{2,\omega,p}$ denote the corresponding quantities in medium 1 and 2 respectively for all directions incoming to the interface. As shown in Fig. 2, the normal vector **n** points outward from the interface into the medium 2. Eq. (11) states that the net energy flux in a particular frequency band $\omega$ and polarization $p$ from the interface toward medium 2 is a sum of the reflected energy flux from medium 2 (in the same frequency band $\omega$ but summed over all polarizations) and the transmitted energy flux from medium 1 in the same frequency band $\omega$ (summed over all polarizations of material 1) and travels with velocity $v_{2,\omega,p}$. Based on the effective equilibrium temperature on either side of the interface, this leads to an interfacial thermal resistance in this frequency band of,

$$R = \frac{2\left(\sum_p v_{1,\omega,p} C_{1,\omega,p} + \sum_p v_{2,\omega,p} C_{2,\omega,p}\right)}{\sum_p v_{1,\omega,p} C_{1,\omega,p} * \sum_p v_{2,\omega,p} C_{2,\omega,p}}$$

(12)

The total interfacial thermal resistance may easily be calculated as inverse of the total conductance over the entire frequency range.

*Heat flux and thermal conductivity*

The net heat flux across any surface $f$ may be calculated as

$$Q_f = \sum_{p,nbands} \int v_{\omega,p} e^{"}_{f,\omega,p} \mathbf{s}.\mathbf{n} d\Omega$$

(13)

Here, **n** is the normal vector pointing outward from the surface into the domain. For a thermal gradient in the x direction, the net heat transfer rate $Q$ and the corresponding thermal conductivity may computed from Eq. (13) by considering transport across any vertical line in the domain.

## III. PHONON DISPERSION AND SCATTERING RATES

For simulating mode and frequency-dependent effects in phonon transport across interfaces, we choose the particular case of silicon. The phonon dispersion for Si is computed using the environment-dependent interatomic potential (EDIP) [44]. The Harrison potential [45] is used to calculate the phonon dispersion of germanium.

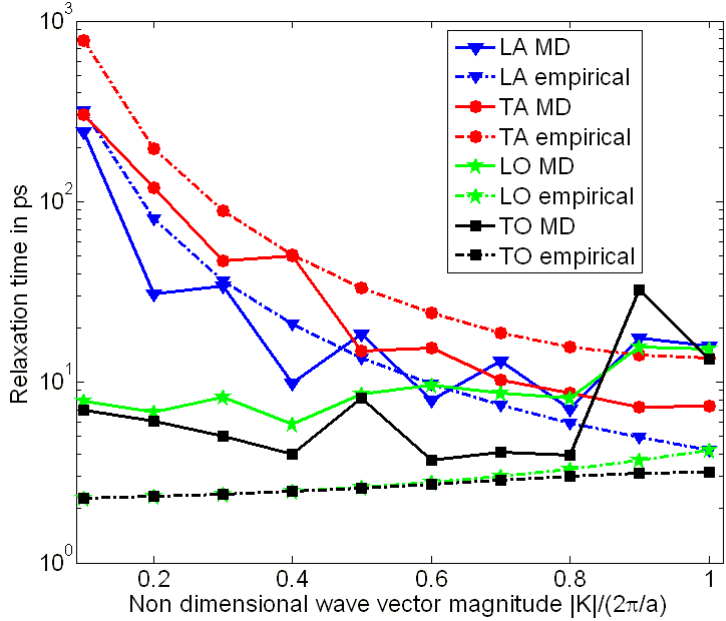

*Fig. 3: Comparison of empirically calculated relaxation time using Eq (15) vs. those computed from molecular dynamics in [48] for phonons in [100] direction of silicon at 300 K*

Most simulations presented here are for a superlattice period length of 10 nm or greater. We therefore assume phonon dispersion and scattering rates to be the same as those for the corresponding bulk material [38]. Several forms for the functional dependency of scattering rates on frequency and temperature for different polarizations exist. The simplest approximation is that of Callaway where the scattering rates were derived from a first-order perturbation expansion [46]. Here we use Klemens' approximation for impurity scattering rates and an Umklapp scattering rate of the form used in [12],

$$\begin{aligned}\tau_{im}^{-1} &= A\omega^4 \\ \tau_u^{-1} &= BT\omega^2 e^{-C/T}\end{aligned} \quad (14)$$

where $A = 1.32 \times 10^{-45}$ s$^3$, $B = 1.73 \times 10^{-19}$ s/K and $C = 137.39$ K for Si. For Ge, the parameters fitted are $A = 2.4 \times 10^{-44}$ s$^3$, $B = 3.35 \times 10^{-19}$ s/K and $C = 57.6$ K. These constants are determined by fitting the bulk thermal conductivity predicted by the model to experimental values in [47]. Comparison against the experimental data of Glassbrenner and Slack is excellent for the entire temperature range for the bulk thermal conductivity of both Si and Ge computed using these relaxation times. Fig. 3 shows a comparison of the relaxation time of phonons in Si due to Umklapp scattering at 300 K computed using Eq. (14) against those derived by Henry and Chen [48] using molecular dynamics simulations. Although MD results give a relaxation time that includes anharmonicity to all orders, Umklapp scattering (*i.e.*, third-order anharmonicity) is expected to be strongly dominant at $T$ = 300K [46]. Agreement between the MD results and our empirical formulae for relaxation time for Si is fair for acoustic phonons but unsatisfactory for optical phonons throughout the entire Brillouin zone. Part of the discrepancy may also be attributable to limitations of classical MD in predicting the relaxation times [49].

## IV. NUMERICAL METHOD

The Boltzmann transport equation for the energy density of phonons is solved using the finite volume method described in [50] with a third-order spatial discretization and a superbee flux limiter [51] to control spatial oscillations. A uniform structured mesh has been used for all simulations with mesh sizes varying from 100 – 400 cells in the *x* direction depending on the domain length. A uniform angular discretization of 8x8 control angles in the octant is used. Phonon dispersion curves are generated by sampling 1000 points in the wave vector space in the [100] direction of both Si and Ge. This detailed band structure is then divided into 26 coarser non-uniformly spaced frequency bands. The specific heat, ballistic thermal conductance and thermal conductivity contributed by each band is accurately calculated using a finer discretization within the band for integration purposes. Subsequently, the group velocity and scattering rate of each coarse band is calculated from its ballistic thermal conductance and thermal conductivity as,

$$v_{g,\omega,p} = \frac{\frac{\partial}{\partial T}\left(\int_{\Delta\omega} \frac{\hbar\omega * v_{g\omega,p}}{e^{\hbar\omega/k_bT}-1} D_p(\omega)d\omega\right)}{\frac{\partial}{\partial T}\left(\int_{\Delta\omega} \frac{\hbar\omega}{e^{\hbar\omega/k_bT}-1} D_p(\omega)d\omega\right)} \qquad (15)$$

$$(v_g\tau)_{\omega,p} = \frac{\frac{\partial}{\partial T}\left(\int_{\omega=0}^{\omega_{\max}} \frac{\hbar\omega * v_{g\omega,p}^2 * \tau_{\omega,p}}{e^{\hbar\omega/k_bT}-1} D_p(\omega)d\omega\right)}{\frac{\partial}{\partial T}\left(\int_{\omega=0}^{\omega_{\max}} \frac{\hbar\omega * v_{g\omega,p}}{e^{\hbar\omega/k_bT}-1} D_p(\omega)d\omega\right)} \qquad (16)$$

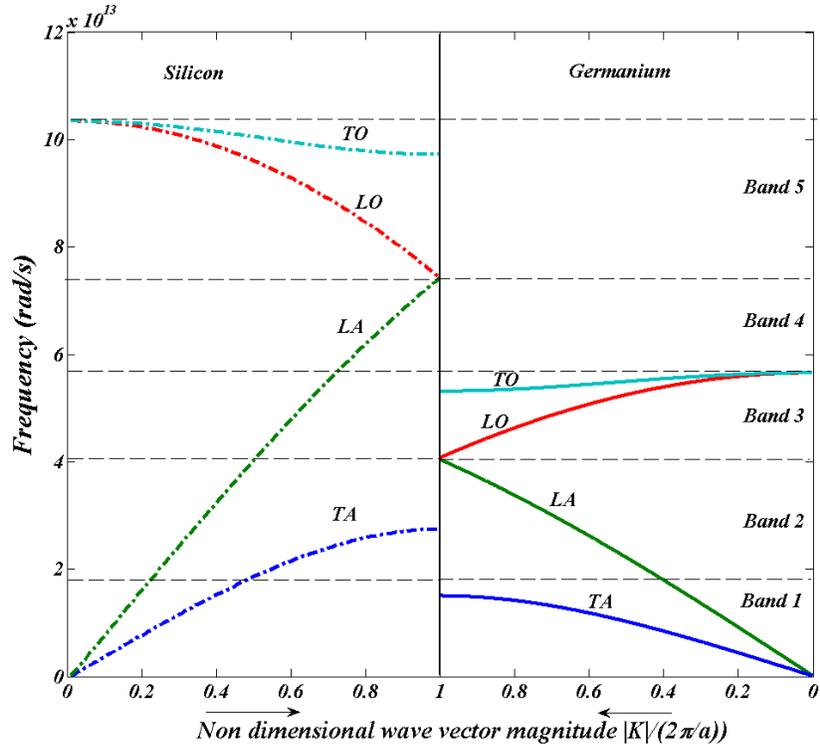

*Fig. 4(a): Phonon dispersion in Si and Ge along [100] direction*

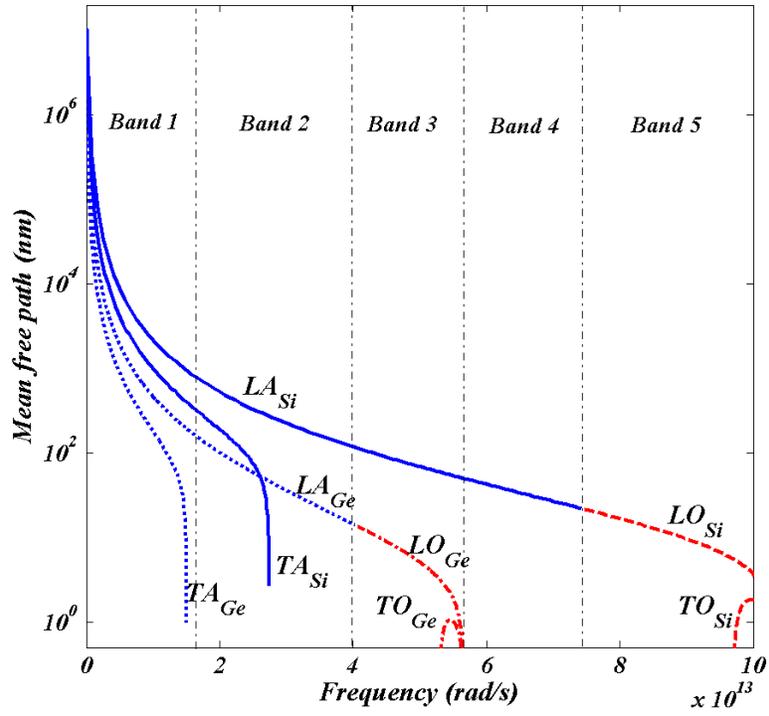

*Fig. 4(b): Mean free path of phonons in Si and Ge*

## V. RESULTS AND DISCUSSION

All simulations presented here are performed with thermophysical properties calculated at 300 K. At this temperature, the dominant angular frequency of phonons, computed using $\hbar\omega \sim 2.8 k_b T$ (computed using a linear dispersion with a spherical k-space) is 2.3 THz. A natural classification in quantifying heat flow through nanostructures is the division of phonons into acoustic and optical branches. The two transverse branches of phonons in Si and Ge in the [100] direction are degenerate. To better interpret results and to delineate the role of each phonon group, we further classify the phonon spectrum into five discrete bands as shown in Fig. 4(a). Fig. 4(b) shows the frequency dependence of mean free paths for different phonon groups in both Si and Ge. A brief description of the five discrete bands is given below in Table I. We emphasize that these 5 bands are identified for post-processing purposes only; the BTE computation itself employs a much finer discretization. It is also instructive to analyze the effective 'temperature' of the optical and acoustic phonons to understand the non-equilibrium that exists between the lattice bath and different phonon groups.

|  | Band 1 | Band 2 | Band 3 | Band 4 | Band 5 |
|---|---|---|---|---|---|
| *Description* | LA, TA in Si; LA, TA in Ge | LA, TA in Si; LA in Ge | LA in Si; LO, TO in Ge | LA in Si; None in Ge | LO, TO in Si; None in Ge |
| *Frequency range (rad/s)* | $(0\text{-}1.5)\times 10^{13}$ | $(1.5\text{-}4.1)\times 10^{13}$ | $(4.1\text{-}5.7)\times 10^{13}$ | $(5.7\text{-}7.4)\times 10^{13}$ | $(7.4\text{-}10)\times 10^{13}$ |
| *% of total specific heat (Si)* | LA < 1 % TA = 2.7 % | LA = 2.34 % TA = 39.1 % | LA = 4.62 % No TA band | LA = 10.8 % No TA band | LO = 14.7 % TO = 26.2 % |
| *% of total specific heat (Ge)* | LA < 1 % TA = 35.8 % | LA = 16.4 % No TA band | LO = 16.1 % TO = 31 % | No phonons | No phonons |
| *Average group velocity $v_g$ (ms$^{-1}$) in Si* | LA: 7181 TA: 3206 | LA: 6842 TA: 1325 | LA: 6272 No TA band | LA: 5401 No TA band | LO: 3802 TO: 501 |
| *Average group velocity $v_g$ (ms$^{-1}$) in Ge* | LA: 4013 TA: 830 | LA: 3271 No TA band | LO: 2130 TO: 291 | No phonons | No phonons |

*Table I: Description of the discrete phonon bands shown in Fig. 4(a). The abbreviations LA and TA represent longitudinal and transverse acoustic phonons respectively. LO and TO represent longitudinal and acoustic optical phonons.*

The "temperature" of acoustic and optical phonons, $T_{AP}$ and $T_{OP}$, may be calculated in the same manner as the lattice temperature, except now the sum in Eq. (3) runs over only acoustic or optical bands. It is important to emphasize that this mode-wise 'temperature' is not the thermodynamic temperature; it is proportional to the energy contained in the corresponding phonon group and is useful in interpreting results.

*V.I. Transport across a single interface*

We perform a systematic calculation of thermal conduction normal to a single interface between Si and Ge, as shown in Fig. 1(a). The Si and Ge portions are of equal length *L* and may be conceived of as connected to bulk contacts at given temperatures on either ends. Figure 5 shows the dependence of heat flux (for $T_1$-$T_2$=1 K) through the domain as a function of *L* (This is also the conductance between the contacts). As expected, the heat flux decreases as the domain length *L* increases because of increased phonon scattering in the bulk region of both Si and Ge (the behavior should approach $L^{-1}$ at large *L*). However, as *L*→ 0, the heat flux must become a constant in the ballistic limit where phonons do not interact with each other in the bulk, and only scatter from interfaces. It can be seen that even in the region *L*<50 nm in Fig. 5, there is a strong signature of diffusive behavior. The effective mean free paths (mfp) of phonons (equivalent to a gray approximation) in Si and Ge at 300 K are approximately 270 nm and 200 nm respectively. Thus, if a gray treatment were used, *L*<50 nm would be expected to result in nearly ballistic behavior. We would expect to find an approximately constant resistance equal to the sum of the ballistic thermal resistance at the two contacts and the interfacial thermal resistance. Furthermore, in the ballistic limit, the temperature drop would be limited to the interfaces and contacts, and the temperature would not vary in the bulk. Neither behavior is seen in our calculations, suggesting that phonon bands with very small mean free paths contribute substantially to overall resistance. The specific mechanisms for this behavior are discussed below.

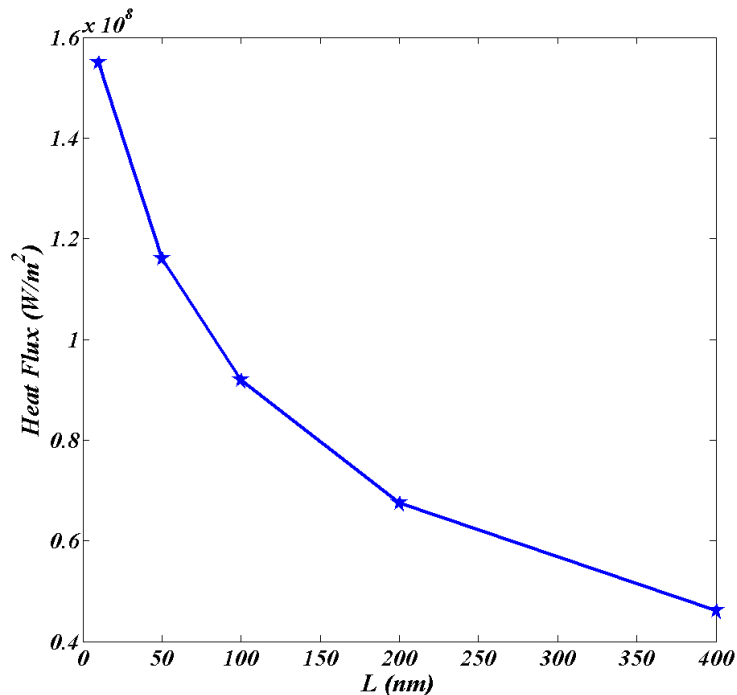

*Fig. 5: Heat flux across the domain for varying L*

*Temperature Profile:* Figure 6 shows the computed temperature profile across the domain for three different values L: 10nm, 50 nm and 400 nm. The variable plotted is the dimensionless temperature, defined as $\theta=(T-T_2)/(T_1-T_2)$ versus $x^*$ $(=x/2L)$. For each $L$, three temperatures are plotted corresponding to the lattice bath, acoustic phonons and optical phonons, respectively. The position of the interface is $x^* = 0$. We consider thermal behavior in the two extremes of domain length first, and then consider the intermediate length, $L$=50 nm.

*L = 400 nm:* It can be seen from Fig. 6 that at $L$=400 nm (solid lines), there is almost no distinction in the temperature of the acoustic phonons (AP) and the optical phonons (OP), and both are equal to the lattice temperature throughout the domain. This means that when the domain length is large enough there is strong scattering of phonons with each other, and equilibrium exists among different phonon polarizations and bands. There is significant gradient in the bulk of the two solids (with a higher slope in Ge because of smaller phonon mfp) since the mfp of 3-phonon scattering events is smaller than the sample length. A significant temperature drop across the interface between Si and Ge can also be observed and relates to the interfacial thermal resistance across the structure.

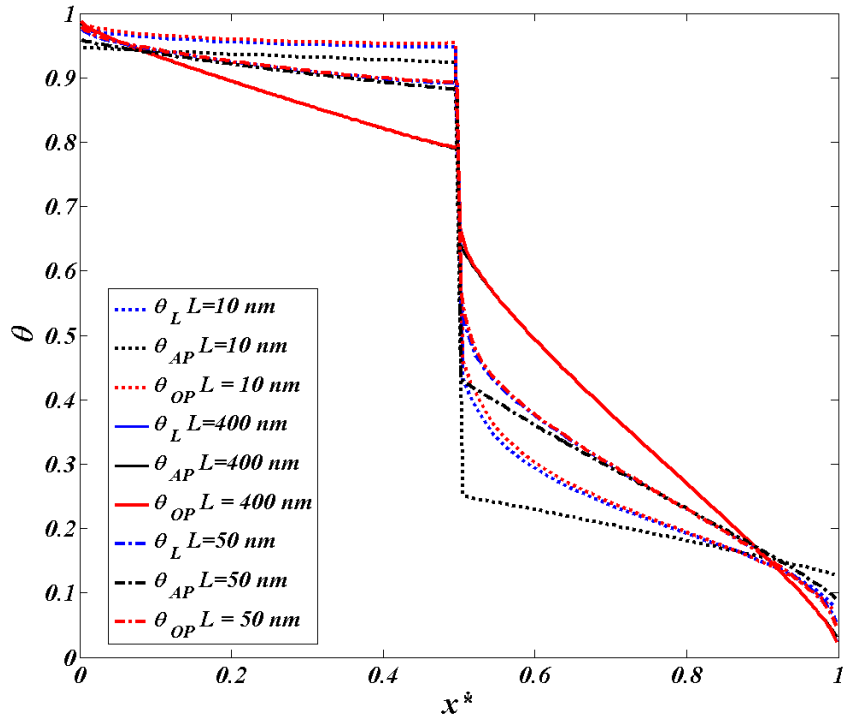

Fig. 6: Lattice, acoustic phonon and optical phonon temperature across the domain for L=10 nm, 50 nm, 400 nm

These predictions match the expected behavior in the diffusive phonon transport regime where Fourier's law is valid.

*L = 10 nm:* As mentioned earlier, $L$ = 10 nm represents a length scale for which quasi-ballistic transport of phonons is expected in both Si and Ge. The primary reason for temperature drop in this regime is due to the interaction of phonons with physical boundaries and interfaces. The dotted lines in Fig. 6 represent the lattice, AP and OP temperatures across the domain. The lattice temperature ($T_L$) is nearly constant in Si and almost equal to the optical phonon temperature, $T_{OP}$. These are also fairly close to the temperature of acoustic phonons, $T_{AP}$. The contribution of a particular phonon band to the determination of the lattice temperature is determined by the value of $C_{\omega,p}/\tau_{\omega,p}$ of that band (see Eq. (3)). $T_L$ is therefore closer to $T_{OP}$ since OP exchange energy with the lattice at much faster rates due to their high scattering rates as compared to AP. OP have much lower mfp (~1-10 nm) and therefore one expects a significant drop to exist in $T_{OP}$ in Si. The reason that $T_{OP}$ is almost constant in the Si region is because Ge does not support any phonon modes in the frequency range of OP in Si (see Fig. 4) and therefore OP in Si do not carry any energy across the interface. The interface represents an impenetrable (fully-reflecting) boundary condition for band 4 and band 5 phonons as these lie above the frequency range of all phonons in Ge. In this case the only mechanism that allows $T_{OP}$ drop in the bulk region of Si is through scattering to AP in bands 1-3 in the bulk. AP in bands 1-3 have long mfp (~100-1000 nm) and therefore do not interact sufficiently with the lattice to pick up energy carried by the optical modes. Therefore the temperature of OP remains almost constant through the Si region of the domain. Similarly $T_{AP}$ is also disproportionately determined by the band 4 LA phonons in Si as these have high values of $C_{\omega,p}/\tau_{\omega,p}$, leading to a constant $T_{AP}$ in the Si region.

A sharp drop in $T_L$ across the interface (~50% of the total temperature difference) is apparent for $L$ = 10 nm, and the drop is even more pronounced for $T_{AP}$. Furthermore, a large difference between $T_L$ (~$T_{OP}$) and $T_{AP}$ exists near the interface on the Ge side. The reason for this behavior relate to the interactions between bands on opposing sides of the interface. Band 3 AP phonons in Si correspond in frequency to band 3 OP phonons in Ge (see Fig. 4(a)); there is no direct transmission to AP in Ge for this band. As a result the interface offers a high resistance to AP originating in Si. A non-equilibrium region is created in Ge near the interface due to the selective transport of energy from band 3 AP in Si to slow-moving and highly scattering OP of Ge. Furthermore, band 1 and band 2 phonons in Ge also have high mean free path and do not scatter efficiently with OP. These factors combine to render the transport in Ge quite diffusive even at $L$=10 nm. For the same reason, the heat flux shown in Fig. 5 does not tend to become a constant even at low values of $L$. An interesting feature of these complex energy interactions at the interface is that OP in Ge play an active role in offering thermal resistance. This is contrary to the view frequently adopted in the literature that optical phonons do not play an important role in heat conduction.

*L = 50 nm:* This length scale represents a regime between the two extremes discussed above, and the same mechanisms apply here. While a non-equilibrium condition exists between AP and the lattice near the interface on the Ge side, equilibrium is restored near the contacts since band 2 acoustic phonons in Ge tend to scatter more efficiently with the lattice. In Si, for similar reasons as discussed for *L*=10 nm, AP, OP and lattice are in equilibrium, and there is a negligible drop across the bulk region. The temperature drop across the interface is significant, and contributes to about 40% of the overall temperature drop between the boundaries.

*Heat Flux:* Figures 7 and 8 show the fractional values of heat flux across the domain carried by the different phonon groups shown in Fig. 4(a). The relative heat flux carried by band *i* is the ratio of heat flux carried by band *i* to the total heat flux calculated by Eq. (13). The heat flux for band *i* across any face may be calculated in the same way as Eq. (13) except that the summation is over frequencies contained in band *i* and not over all frequency bands. (The sum of the relative heat flux over all bands must be unity at each point in the domain).

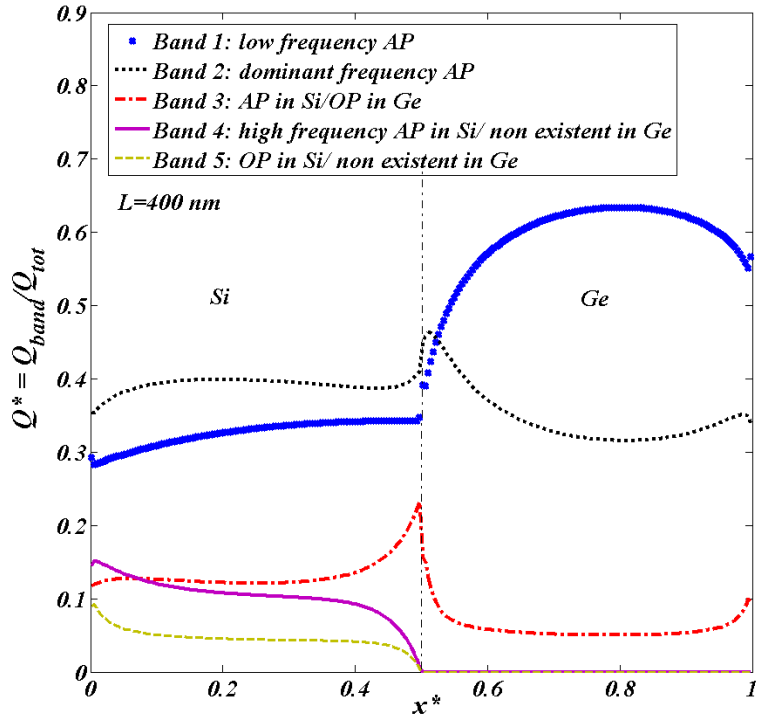

*Fig. 7: Fractional contribution to heat flux of various phonon groups for L=400 nm. x\*=0.5 represents the position of the interface between Si and Ge*

*L=400 nm*: The heat flux profiles for this case are shown in Fig. 7. Analysis of temperature profiles for this case (Fig. 6) indicates that transport at this length scale is primarily in the diffusive regime in both Si and Ge. We consider the Si and Ge regions in turn.

On the silicon side, we see that the main contributors to heat flux are bands 1 and 2. The OP contribution to the heat flux (bands 4 and 5) in the Si region is about 5%. The energy from these bands is redistributed to bands 2 and 3 in a narrow region near the interface on the Si side; these bands carry energy over to the Ge side. On the Ge side, band 2 (AP) and band 3 (OP) receive energy from bands 2 and 3 on the Si side. Both bands scatter energy to band 1 (AP) phonons in the Ge region. The bulk of the transport on the Ge side is due to band 1 and band 2 (AP) phonons. In the attempt to categorize heat flow through the somewhat coarse discretization into five bands, it seems like low frequency modes in Ge are the main contributors to thermal transport. It is also noteworthy that strong changes in heat flux profile in both materials exist only near the interface and boundaries, while the relative contribution is almost constant in the bulk. This is in keeping with the near-equilibrium nature of transport for large $L$, and the fact that $T_L$, $T_{AP}$ and $T_{OP}$ are nearly identical in Fig. 6. Thus, scattering to other bands is relatively small in the bulk, keeping the heat flux in each band nearly constant; the bands tend to act as parallel channels in the bulk. The only departures from equilibrium occur due to the influence of interfaces and boundaries.

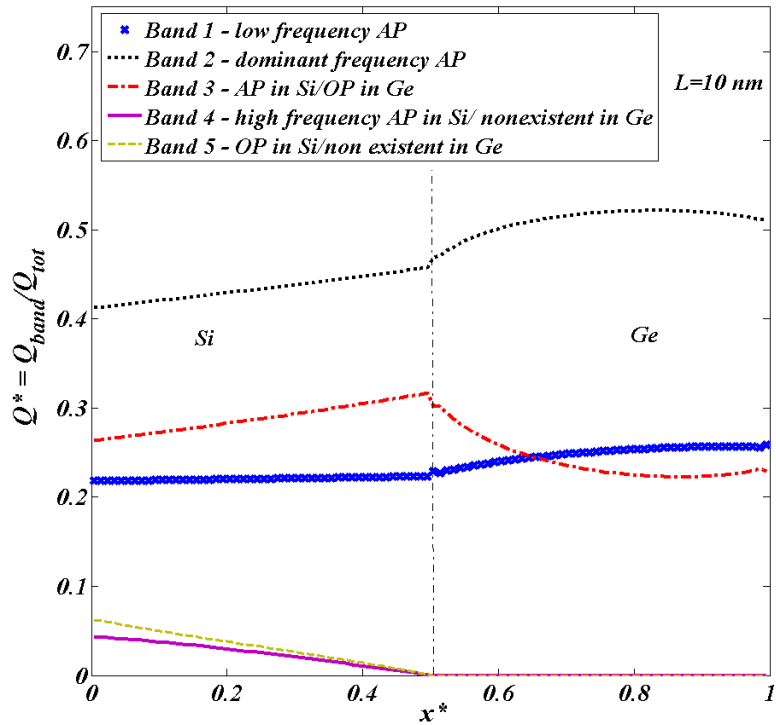

*Fig. 8: Fractional contribution to heat flux of various phonon groups for L=10 nm. x\*=0.5 represents the position of the interface between Si and Ge*

$L=10$ nm: We again consider the two sides in turn. From Fig. 8 it can be seen that in the Si region the major contributor to heat flux are band 2 phonons (corresponding to the dominant phonon frequency) because of their high ballistic thermal conductance at 300 K. Band 3 and band 1 phonons also contribute significantly to heat flux in the domain. Bands 4 and 5 phonons together contribute only about 10% of heat flux near the Si left boundary but their contribution quickly decreases to zero near the Si/Ge interface since none of the energy in this frequency range can be carried across the interface. The heat flux in these modes is mostly transferred to band 2 and band 3 phonons, as seen in Fig. 8. Band 1 phonons have very long mean free paths (a few microns) and do not interact with the lattice; therefore the heat flux of band 1 phonons is constant in the Si region.

We now turn to the Ge side in Fig. 8. Band 4 and 5 phonons do not exist on the Ge side, and therefore carry no heat flux. Band 1 phonons are AP phonons with long mean free path, and travel nearly ballistically through the region, albeit picking up some energy from band 3. Significant energy from band 3 AP in Si is transferred to band 3 OP in Ge at the interface. This energy is scattered in part to band 2 in Ge, a small amount to band 1 and the rest remains in band 3, as seen in Fig. 8. Despite this redistribution, OP in Ge still contribute about 25% of the total heat flux. Overall, band 2 phonons dominate heat flux across the structure as they lie near the dominant frequency range, have high ballistic conductance, and are matched in the frequency spectrum on the two sides of the interface.

*V.II. Transport in superlattice structures:*

We now turn our attention to understanding the role of mode- and frequency-dependent phonon transport in realistic bulk nanostructures. In particular, we analyze heat conduction in the cross-plane direction of a planar 1D superlattice structure shown in Fig. 1(b). While the period length of these superlattices is in the nanometer range, a bulk superlattice structure contains many such periods and exhibits diffusive behavior. The important parameter to predict in this case is the effective thermal conductivity. We apply the periodic jump boundary condition detailed in Eq. (9). This condition has been applied to gray phonon transport in superlattices and nanocomposites [37]. The parameters of interest are the superlattice period length $L_P$ and the volume fraction of Si and Ge in the superlattice. The volume fraction of Si is $\varphi = L_{Si}/L_P$. $L_P$ is inversely proportional to the density of interfaces between Si and Ge. Mode- and frequency-dependent phonon heat conduction is simulated for various values of $\varphi$ and $L_p$ to understand the ballistic diffusive transition and effects of volume fraction.

*Temperature and Heat Flux Profiles*: To illustrate the general trends we choose the case of $\varphi=0.5$. Fig. 9 shows the temperature profiles ($T_L$, $T_{AP}$, $T_{OP}$) across the unit cell of the superlattice for two different values of period length $L_P = 20$ nm and 1000 nm. Because the volume fraction is 0.5, the superlattice is created by stacking equal lengths of Si and Ge.

For $L_p = 1000$ nm, the values of $T_L$, $T_{AP}$ and $T_{OP}$ are almost equal to each other throughout the domain indicating that phonons in different bands are in equilibrium with each other across the length of the superlattice. This means that there is no net transfer of energy from acoustic bands to optical or vice versa. The slight non-equilibrium resulting from transfer of energy from AP in Si to OP in Ge in the band 3 frequency range is confined to a very small region (compared to $L_P$) near the interface. Although temperature profiles generally exhibit a diffusive character, temperature jumps across the interfaces still contribute to about 20% of the total temperature drop. This signifies that $L_P$ is not large enough that transport is primarily determined by bulk scattering of phonons.

The dash-dot lines in Fig. 9 correspond to $T_L$, $T_{AP}$ and $T_{OP}$ for $L_P = 20$ nm. Even in this case, the average AP and OP temperatures in Si are close to each other, implying that different phonon groups are in equilibrium with each other. The reason that such equilibrium exists at this small length scale is the same as for transport across a single interface. As before, band 4 and band 5 phonons in Si have no counterpart on the Ge side because of the mismatch in the spectrum (Fig. 4(a)). As a result, the two Si/Ge interfaces behave as purely reflecting boundaries, leading to a flat temperature profile for the OP mode. Since the OP mode has a small relaxation time (low mfp), it contributes disproportionately to the determination of $T_L$ (see Eq. 3). Furthermore, $T_{AP}$ too is determined disproportionately by band 4 high-frequency phonons which have short relaxation times, equilibrating it with the lattice. As a result $T_{OP}$, $T_{AP}$, and $T_L$ are nearly identical in silicon. However, we should note that although $T_{AP}$ and $T_{OP}$ are nearly equal in Si, the low and moderate frequency acoustic modes in Si are in non-equilibrium with the lattice. On the Ge side however, the optical modes receive a considerable energy flux from band 3 phonons in Si and a marked difference between $T_{AP}$ and $T_{OP}$ exists. Unlike the case of the 1000 nm period length, AP in Ge in this case are not able to interact with the optical phonons because of the small length scales, and the non-equilibrium between these two groups exists over the entire region in Ge.

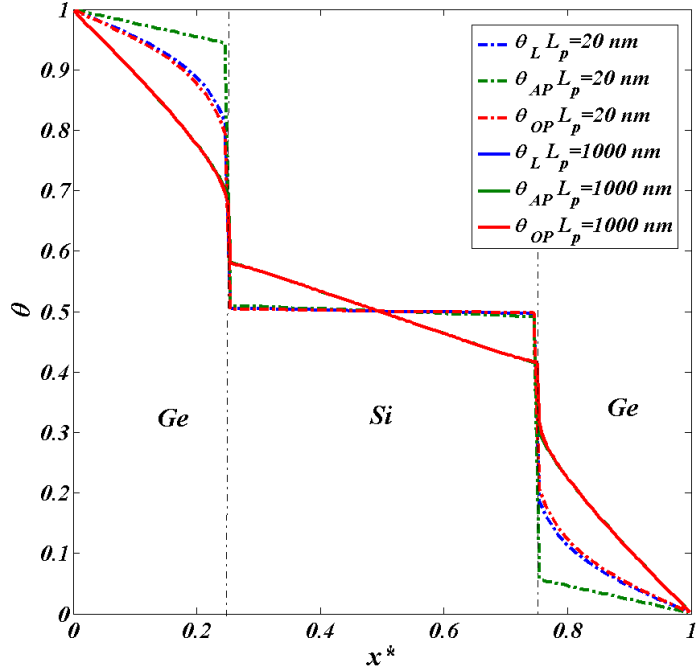

*Fig. 9: Lattice, acoustic phonon and optical phonon temperature across the superlattice unit cell for $L_P$=20 nm, 1000 nm. $\varphi$=0.5*

Figure 10 shows the fractional heat flux of the acoustic and optical modes throughout the domain (not the bandwise description as adopted in the previous section) for $L_{Si}$ = 10 nm, 60 nm and 1000 nm at $\varphi$=0.5 ($L_P$=2*$L_{Si}$). Clearly, as explained above, optical phonons carry no heat flux in Si at small length scales because of spectral mismatch. However, their contribution increases as $L_P$ is increased since acoustic phonons in Si now have sufficient residence time to scatter some energy to optical phonons. Most transport in Si is through acoustic phonons, again because these modes can match the Ge spectrum better.

The picture in Ge is somewhat different, as seen in Fig. 10. Acoustic phonons are still the dominant carrier at all length scales. However, optical phonons play more of a role in transporting heat in Ge, particularly at small length scales. OP in Ge receive energy from band 3 AP of Si (see Fig. 5) at the interface. At small length scales, scattering of energy from OP to AP in Ge cannot occur. Consequently, OP continue to carry energy through the bulk of the Ge slab, contributing as much as 25% to the total heat flux. Their contribution progressively decreases as $L_P$ increases because scattering to AP becomes more pronounced.

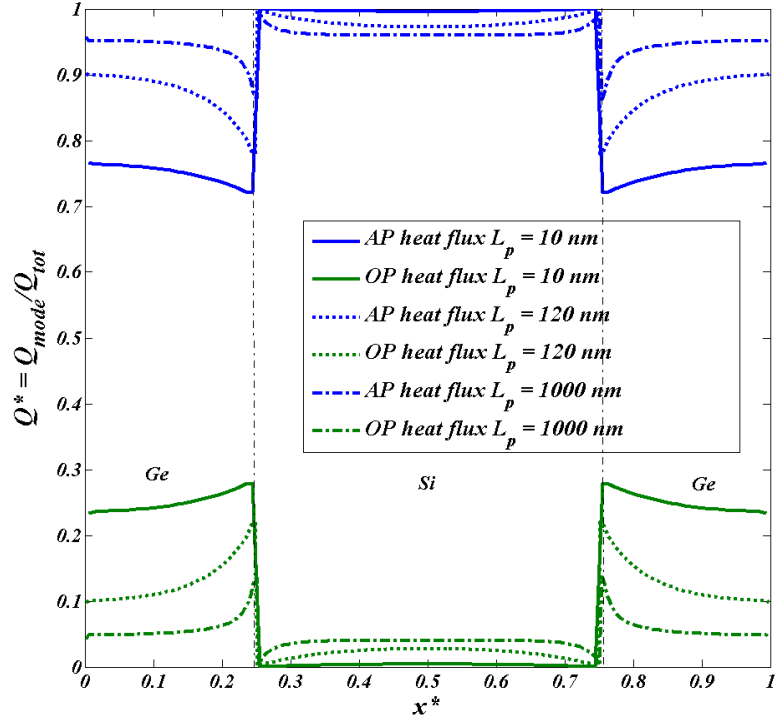

*Fig. 10: Fractional heat flux vs non dimensional position for $L_{Si}$ = 10 nm, 60 nm and 100 nm at $\varphi=0.5$*

*Thermal Conductivity:* Fig. 11 shows a plot of the effective thermal conductivity of the superlattice for various volume fractions and different superlattice period lengths. The values of thermal conductivity predicted by the model result from a competition between interface and volumetric scattering. For period lengths < 50 nm, the thermal conductivity is only a function of the period length $L_P$ (or the physical interface density) and is independent of the volume fraction of Si. This is borne out by the fact that at 300 K, the bulk thermal conductivity of Si is about 155 W/m/K and that of Ge is about 61 W/m/K. The value of thermal conductivity of $Si_\varphi Ge_{1-\varphi}$ superlattice for $L_P$ <50 nm is less than 10 W/m/K and decreases as $L_p$ decreases. These values suggest that diffuse phonon scattering at the interface causes a strong reduction in the effective thermal conductivity; values much below the thermal conductivity of the constituent materials may be obtained. On the other hand, at large period lengths, bulk scattering in Si and Ge is expected to govern thermal resistance and the effective thermal conductivity should be dependent on the volume fraction of the individual constituents. In keeping with the effective medium approximation (EMA) for a superlattice, the thermal conductivity should simply be a volume average of the thermal conductivity of Si and Ge for $L_p \to \infty$. For $L_P > 1$ μm we indeed see a strong dependence of thermal conductivity on the volume fraction of Si. At $L_P = 1$ μm, the spread in the effective thermal conductivity ranges from approximately 30 W/m/K at $\varphi=0.2$ to 50 W/m/K at $\varphi=0.8$, and the difference grows as the period length is increased. However, the presence of interfacial thermal resistance still results in superlattice thermal conductivity values below that of the constituent materials and that predicted by EMA.

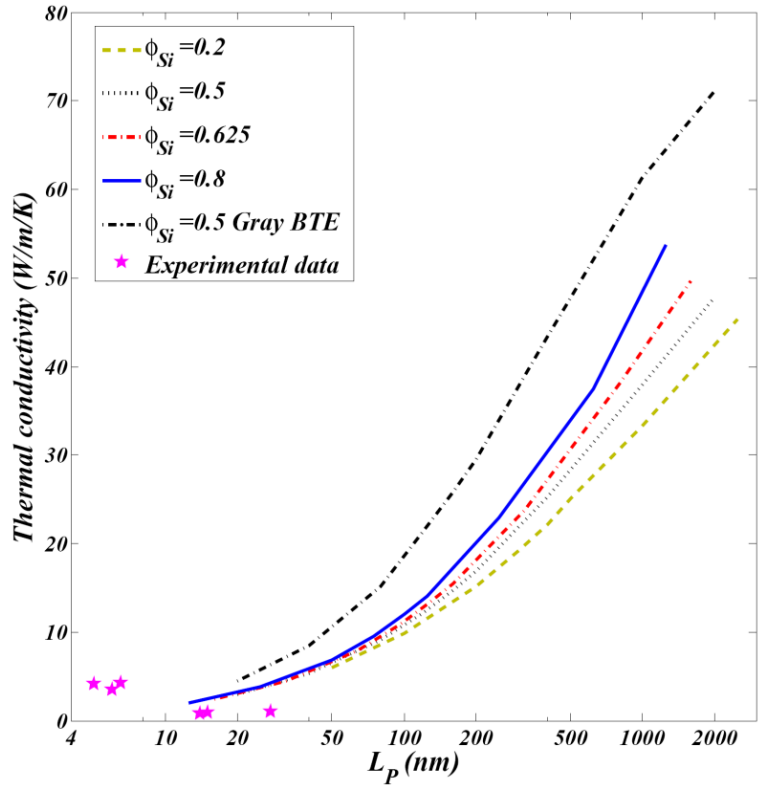

*Fig. 11: Thermal conductivity of superlattice (at 300 K) as a function of the period length at various values of Si volume fraction $\varphi_{Si}$*

The computed values of thermal conductivity are also considerably lower than those predicted using a gray model for phonon transport. The gray values are obtained from an averaging (Eq. (15) and (16)) over the acoustic branches. The parameters used in simulation are shown in Table II. The thermal conductivity based on the non-gray BTE simulations in Fig. 11 implies a thermal boundary resistance of approximately $3.3 \times 10^{-9}$ m$^2$K/W at very low periods. This value is higher than the full dispersion DMM resistance of $2.72 \times 10^{-9}$ m$^2$K/W (using Eq. (12)) and also much greater than the gray value of $2.09 \times 10^{-9}$ m$^2$K/W computed using the parameters in Table II. This additional thermal resistance compared to a simple DMM estimate comes from the energy transfer from optical to acoustic phonons in Ge. This phenomenon is very similar to the additional thermal resistance encountered at a metal dielectric interface due to relaxation of thermal energy from electrons to phonons near the interface in the metal [52, 53]. The gray BTE results depart significantly from the non-gray calculations at period lengths > 100 nm (>50% at 1000 nm). The reason for such behavior is the fact that low lying acoustic phonons with very long mean free paths are only impeded by interfaces, suppressing the heat flux significantly. This phenomenon is not captured by the gray model.

Previously published gray simulations [30] of superlattice thermal conductivity predicted an interface thermal resistance value of $3.4 \times 10^{-9}$ m$^2$K/W, close to the non-gray value computed here. However, we believe that this similarity is only fortuitous. The parameters used for the gray model in [30] were estimated using the sine curve approximation for phonon dispersion which underestimates the group velocity of LA phonons near the zone center and at the ends, and the computed thermal interfacial resistance is a result of this specific approximation.

|  | Si | Ge |
|---|---|---|
| *Specific heat (J/m$^3$/K)* | $0.95 \times 10^6$ | $0.92 \times 10^6$ |
| *Group velocity $v_g$ (ms$^{-1}$)* | 2759 | 1627 |
| *Relaxation time(s)* | $6.26 \times 10^{-11}$ | $7.23 \times 10^{-11}$ |
| *Interfacial thermal resistance (m$^2$K/W )* | $2.09 \times 10^{-9}$ | |

*Table II: Gray model parameters for Si and Ge. Only the acoustic branches have been used for obtaining the gray values*

For comparison we also show data points from the experimental results of Lee et al [29] in Fig. 11 at 300 K. The experimental data in the 10-30 nm period length exhibit lower thermal conductivity than the computational results of the present work. The trend from the current predictions leads to lower thermal conductivity than the experimental data for low period length (<10 nm). This may point to deficiencies in the DMM in overpredicting interfacial resistance at epitaxial Si/Ge interfaces. Nevertheless, results from the current work, may help explain the high measured thermal resistance of several other acoustically mismatched interfaces [1] where lattice dynamical calculations have predicted thermal resistance that is significantly lower than the measured values. Our computations reveal that scattering between optical and acoustic modes close to the interface could play a significant role in interfacial thermal resistance, an effect not captured by lattice dynamics models.

## VI. CONCLUSIONS

In this paper, we present simulations of non-gray phonon transport across an interface between Si/Ge. We clearly delineate the role of each phonon group in thermal transport. Furthermore, the transition from the quasi-ballistic to the diffusive regime is studied. The non-equilibrium that results between the optical and acoustic phonons near the interface is analyzed and explained. Transport across a single rough interface is used as the basis to analyze in detail the phonon transport in periodic superlattice structures. We calculate the thermal conductivity of Si/Ge superlattices as a function of the superlattice period and the volume fraction of the constituent materials. It is found that at nanoscale period lengths thermal conductivity is a function of the interfacial density and remains independent of volume fraction. Though the EMA limit is not achieved for the period lengths considered here, a strong dependence on volume fraction is seen for superlattice periods of a micron or higher. We also show that a non-gray model of phonon transport leads to a higher interfacial thermal resistance than would be calculated by a gray model, indicating that the mismatch of the phonon spectra between Si and Ge and bulk scattering near the interface play a critical role in determining interface resistance.

All the predictions in this paper have been made under the diffuse mismatch model of phonon transmissivity and anharmonic effects, if any, are not accounted for in the transmissivity. It remains to be seen what the effect of different models of phonon transmissivity and temperature would have on the bulk thermal conductivity of superlattices and nanocomposites. Furthermore, the analysis proposed in this study may be extended to study in detail the thermal transport across metal/dielectric and other dissimilar solid interfaces.

**ACKNOWLEDGEMENTS**

J. Murthy and D. Singh wish to acknowledge use of Purdue's nanoHUB (http://www.nanohub.org) and support from the DARPA-funded IMPACT Center at the University of Illinois, Urbana-Champaign. Support of J. Murthy by the Department of Energy (National Nuclear Security Administration) under award no: DE-FC52-08NA2861 is also acknowledged.

**NOMENCLATURE**:

| | |
|---|---|
| $C$ | gray volumetric specific heat of phonons (Jm$^{-3}$K$^{-1}$) |
| $C_{i,\omega,p}$ | frequency and polarization dependent phonon specific heat in material $i$ |
| $D_{pi}(\omega)$ | phonon density of states of material $i$ |
| $e''$ | directional net phonon energy density (Jm$^{-3}$sr$^{-1}$) |
| $e^0$ | angular average of directional net phonon energy density (Jm$^{-3}$sr$^{-1}$) |
| $e''_{i,\omega,p}$ | directional net energy density of phonons in polarization $p$ and frequency band $\omega$ in material $i$ (Jm$^{-3}$sr$^{-1}$) |
| $e''_{f,\omega,p}$ | directional net energy density of phonons in polarization $p$ and frequency band $\omega$ at the interface (Jm$^{-3}$sr$^{-1}$) |
| $e^0_{\omega,p}$ | equilibrium energy density of phonons in polarization $p$ and frequency band $\omega$ (Jm$^{-3}$sr$^{-1}$) |
| $f$ | non-equilibrium distribution function of phonons |
| $\hbar$ | reduced Planck constant (Js) |
| $k_b$ | Boltzmann constant (m$^2$kgs$^{-2}$K$^{-1}$) |
| $|K|$ | wave vector magnitude (m$^{-1}$) |
| $Kn$ | Knudsen number |
| $L$ | Si/Ge region thickness (single interface domain) (m) |
| $L_{Si}$ | silicon layer thickness in superlattice (m) |
| $L_P$ | superlattice period length (m) |
| $LA$ | longitudinal acoustic |
| $LO$ | longitudinal optical |
| **n** | unit normal vector to the surface |
| $nbands$ | total discrete frequency bands |
| $p$ | phonon polarization index |
| $Q_f$ | heat flux across surface $f$ (Wm$^{-2}$) |
| **r** | position vector (m) |
| $r_{ij}$ | phonon reflectivity at interface from material $i$ to material $j$ |
| $R$ | interfacial thermal resistance (m$^2$KW$^{-1}$) |
| **t** | heat flow direction |
| $t_{ij}$ | phonon transmissivity from material $i$ to material $j$ |
| $T_L$ | lattice temperature (K) |

| $T_{\omega,p}$ | temperature of phonons in frequency band of frequency $\omega$ and polarization $p$ |
| --- | --- |
| TA | tranverse acoustic |
| TO | transverse optical |
| $T_{AP}$ | acoustic phonon temperature (K) |
| $T_{OP}$ | optical phonon temperature (K) |
| $\Delta T_{drop}$ | temperature drop across the unit cell in the heat flow direction (K) |
| $v_{gi}$ | gray phonon group velocity in material $i$ (ms$^{-1}$) |
| $v_{i,\omega,p}$ | frequency and polarization dependent phonon group velocity in material $i$ (ms$^{-1}$) |
| $y$ | y coordinate (m) |

**Greek Symbols**

| $\theta$ | dimensionless temperature |
| --- | --- |
| $\theta_{AP}$ | dimensionless acoustic phonon temperature |
| $\theta_{OP}$ | dimensionless optical phonon temperature |
| $\varphi$ | volume fraction of silicon in the superlattice |
| $\tau_i$ | relaxation time of phonons in material $i$ (s) |
| $\tau_{i,\omega,p}$ | single mode relaxation time of phonons of frequency $\omega$ and polarization $p$ in material $i$ (s) |
| $\tau_{im}^{-1}$ | impurity scattering rate (s$^{-1}$) |
| $\tau_u^{-1}$ | Umklapp scattering rate (s$^{-1}$) |
| $\omega$ | phonon frequency (rads$^{-1}$) |
| $\omega^*$ | discrete phonon frequency bands (rads$^{-1}$) |
| $\omega_{max,i}$ | maximum phonon frequency in material $i$ (rads$^{-1}$) |
| $d\Omega$ | incremental solid angle (sr) |